\documentclass[pre,amsmath,amssymb,floatfix,twocolumn]{revtex4-1}
\usepackage{color}
\usepackage{graphicx,epsfig}
\usepackage[hypertex]{hyperref}
\usepackage{srctex}

\DeclareMathOperator{\ssL}{\scriptscriptstyle{\mathrm{L}}}
\DeclareMathOperator{\ssR}{\scriptscriptstyle{\mathrm{R}}}
\DeclareMathOperator{\LR}{\scriptscriptstyle{\mathrm{(LR)}}}
\DeclareMathOperator{\RL}{\scriptscriptstyle{\mathrm{(RL)}}}

\begin{document}

\date{\today }

\title{Universality and quantization of the power to heat ratio in nano-granular systems} 

\author{N.~M.~Chtchelkatchev}
\affiliation{Institute for High Pressure Physics, Russian Academy of Science, Troitsk 142190, Russia}
\affiliation{L.D. Landau Institute for Theoretical Physics, Russian Academy of Sciences,117940 Moscow, Russia}
\affiliation{Department of Theoretical Physics, Moscow Institute of Physics and Technology, 141700 Moscow, Russia}
\affiliation{Department of Physics and Astronomy, California State University Northridge, Northridge, CA 91330, USA}

\author{A.~Glatz}
\affiliation{Materials Science Division, Argonne National Laboratory, Argonne, Illinois 60439, USA}
\affiliation{Department of Physics, Northern Illinois University, DeKalb, Illinois 60115, USA}

\author{I.~S.~Beloborodov}
\affiliation{Department of Physics and Astronomy, California State University Northridge, Northridge, CA 91330, USA}

\begin{abstract}
We study heating and dissipation effects in granular nanosystems in the regime of weak coupling between the grains. We focus on the cotunneling regime and solve the heat-dissipation problem in an array of grains exactly. We show that the power to heat ratio has a universal quantized value, which is geometrically  protected: it depends only on the number of grains.
\end{abstract}

\pacs{05.60.Gg,05.30.Fk,73.23.Hk,73.63.Rt,05.30.Jp,84.30.Bv}

\maketitle
\section{Introduction}
A major design issue of modern nano-electronic devices at low temperatures is the control of temperature, which first and foremost requires a fundamental understanding of self-generated heat inside these systems.
In particular, {\it overheating} of quantum circuits strongly changes its transport properties --  sometimes irreversibly.
A typical feature of these systems is the non-locality of the heat generation when the charge carriers generate heat in the electrodes in the course of their equilibration inside the junction due to inelastic propagation, the latter being the most difficult to control. The fundamental problem is to understand dissipation mechanisms and to optimize cooling procedures~\cite{giazotto2006opportunities}.

\begin{figure}[h]
  \centering
  \includegraphics[width=\columnwidth]{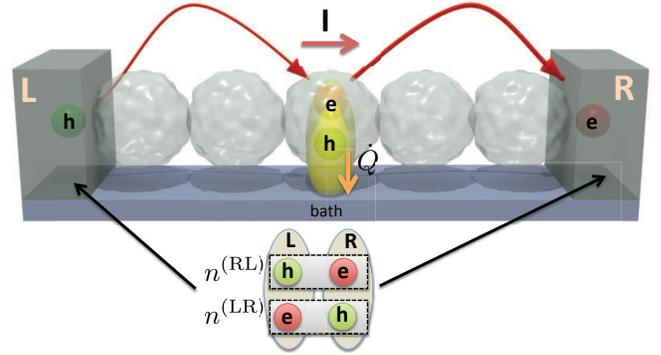}
  \\
  \caption{Sketch of the system under consideration in this work: a chain of weakly coupled nano-grains. The red curved arrows indicate an inelastic cotunneling process leaving behind an electron-hole (e-h) pair in a grain. These processes are responsible for the electron transport in the system ($I$). The energy of this e-h pair is then dissipated into the bath by $\dot Q$.
The form-factor $n^{\LR}_{\Omega>0}$ in Eq.~(\ref{Ig}) can be interpreted as the distribution function of electron-hole pairs left behind by a tunneling event through the system, where the electron sits on the left and the hole on the right lead, respectively. The form-factor $n^{\RL}_{\Omega>0}$ corresponds to the opposite situation.}\label{fig.sys}
\end{figure}

A prototypical system, which is also the most common component of nano-circuits, is a junction made of nanostructured conducting materials weakly coupled to electrodes~\cite{devoret1992single,aleiner2002quantum,beloborodov2007granular} (see Fig.~\ref{fig.sys}). Such systems are the building blocks of promising high-density, high-speed, and low-power memory devices~\cite{rozenberg2004nonvolatile}. The quantum regime in these circuits is realized when \textit{i)} the electrostatic charging energies of the granulars  corresponding to a single charge carrier are much larger than the grain temperatures and bias between the electrodes and \textit{ii)} the bare tunnel resistances exceed the quantum resistance.

In this Letter we study heating of such nano-systems at the example of a chain of nano-crystalline grains or quantum dots in the regime of weak coupling between the grains.
Since we are interested in the low-temperature behavior, we focus on the cotunneling transport regime, which is the dominate transport mechanism in that case.
We solve the heat dissipation problem in this junction exactly and show that the ratio of the total dissipated power $I\cdot V$ to the heat dissipated in the grains, $\dot Q$, has an universal quantized value
\begin{equation}\label{eq1}
  \frac{\dot Q}{I\cdot V}=\frac{n}{n+1}\,,
\end{equation}
where $I$ is the current, $V$ is the bias voltage, and $n$ is an integer corresponding to the total number of grains in the chain.
This rational value in Eq.~\eqref{eq1} is geometrically  protected, i.e., it neither depends on the shape of the grains, nor on the microscopic details of the tunnel junctions.

We assume that all grains are generally different meaning that each grain has its own charging energy, its own shape and its own conducting material (the transparencies of tunnel barriers are also individual). However, we find that the part of heat dissipated in the grain is universal: it is given by a fraction $1/(n+1)$ of the total power $I\cdot V$.

Electrodes are bulk conductors attached to the granulars by the tunnel contacts. We find that each electrode acquires the
universal fraction of heat
\begin{gather}\label{eq2}
  \frac{\dot Q_{\rm electrode}}{I\cdot V}= \frac12 \, \frac{1}{n+1}\,.
\end{gather}
In particular, for a single grain, i.e., $n=1$, half of the power is dissipated in the grain and the other half in the leads.

The physical origin of this quantization is the creation of electron-hole pairs in the grains by a cotunneling electron, which shares part of its energy with those pairs.
The other part of electron energy (the same amount) is dissipated into the leads.

Typically, the current-voltage characteristics, $I(V,T)$, of a granular nanojunction is highly nonlinear both in voltage and temperature. These non-linearities are taken into account when Eqs.~\eqref{eq1}-\eqref{eq2} were derived. We only assumed that the leads are in equilibrium at similar temperatures, energy relaxation processes within each grain are fast enough such that a local equilibrium description is applicable, and that the temperature of the grains are similar to the lead temperatures as well. The latter implies that excess heat from the grains is efficiently transferred into a thermal bath by phonons. A more general situation is discussed at the end of this Letter. In particular, we address
the question of stability of universality and the power to heat quantization, Eqs.~\eqref{eq1}-\eqref{eq2}, with respect to different grain temperatures. We show that the quantization is still valid for biases well exceeding the dispersion of grain and lead temperatures.

The quantization of the power to heat ratio in Eq.~(\ref{eq1}) is a result of inelastic cotunneling processes, which govern the electron transport.
The essence of these processes is that an electron tunnels via virtual states in intermediate grains thus bypassing the huge Coulomb barrier~\cite{devoret1992single,PhysRevLett.78.4482,PhysRevB.58.7882,beloborodov2007granular,PhysRevLett.93.266802,PhysRevB.52.16676,thielmann2005cotunneling,PhysRevLett.95.146806,paul1994cotunneling,PhysRevB.66.045301,PhysRevB.78.075437,PhysRevB.24.2449,huttel2009pumping,dong2005inelastic,PhysRevLett.94.206805}. This can be visualized as coherent superposition of two events: tunneling of an electron into a granule and the simultaneous escape of another electron from the same granule. There are two distinct mechanisms of cotunneling processes, elastic- and inelastic cotunneling. Elastic cotunneling means that the electron that leaves the grain has the same energy as the incoming one. In the case of inelastic cotunneling, the electron coming out of the grain has a different energy than the entering electron (see curved arrows in Fig.~\ref{fig.sys}). This energy difference is absorbed by an electron-hole excitation in the grain, which is left behind in the course of the inelastic cotunneling process. Below we concentrate on the inelastic case, since only this transport mechanism contributes to heating effects. In particular, elastic cotunneling and sequential tunneling do not create electron-hole pairs inside the grain.

At temperatures and voltages below the Coulomb energy, cotunneling typically dominates other mechanisms of electron transport in granular nanojunctions such as the ``the single-charge transistor'' mechanism and sequential tunneling. The single-charge transistor tunneling in a granular device in the quantum case is realized only for rare combinations of parameters and needs in addition well controlled gate voltages applied to each grain. In this case the excess charge carriers can classically stay in the grains for a sufficiently long time~\cite{PhysRevLett.59.109,altshuler1991mesoscopic,devoret1992single,golubev1992charge,likharev1999single,PhysRevB.57.15400,kubala2008violation,shin2011room,prati2012few}. Even then it is difficult to maintain this regime due to charge migration into the nearby gates and insulating areas. The contribution from sequential electron tunneling is exponentially suppressed in the Coulomb blockade regime for temperatures and voltages below the characteristic single-electron charging energy in the grain, $eV$, $T < E_c$.

Each grain is characterized by two energy scales: (i) the mean energy level spacing $\delta$ and (ii) the charging energy $E_c \sim e^2(4\pi\kappa\epsilon_0 a)^{-1}$, where $\kappa$ is the relative static permittivity of the grain material and $a$ is the grain size. We concentrate on the case of metallic grains which are most commonly encountered in applications, where the free electron spectrum can be considered continuous. In this case $E_c$, all involved temperatures, and voltage far exceed $ \delta $. In this regime inelastic cotunneling dominates elastic processes~\cite{devoret1992single,beloborodov2007granular}.

\section{Model calculation}
The current-voltage characteristics, $I(V)$, of the multigranular nanojunction in the cotunneling regime can be found microscopically in several ways starting from the tunnel Hamiltonian, see, e.g., Ref.~\cite{beloborodov2007granular} for a review. We follow the rate-approach, where the current is expressed through the difference of backward and forward electron tunnel rates from one lead to the other, similar to Ref.~\cite{devoret1992single,Korotkov1994JAP,Guinea1998PRB}. To write the rates and observables we use the ``bosonic'' language~\cite{chtchelkatchev2009hierarchical,ChtchelkatchevJPhys2013} that has a direct physical interpretation and allows for analytical calculations. We can write the final result for the current-voltage characteristics, $I(V)$, for an arbitrary number of grains $n$ in the nanosystem as
\begin{equation}\label{Ig}
I = \frac e{\mathcal R}\int_{-\infty}^\infty (n^{\LR}_{\Omega}-n^{\RL}_{\Omega}) P(\Omega)\Omega d\Omega,
\end{equation}
where $\mathcal R=\prod_{i=0}^n R_{i,i+1}/R_q$, $R_q = \pi \hbar/e^2$ is the quantum resistance, and $R_{i,i+1}$ is the bare resistance of the tunnel barrier between the grains ($0<i<n$) or the grains and leads if $i\in \{0,n\}$. $\Omega d\Omega$ is the integration weight over frequencies. Here $n^{\LR}_{\Omega} =\frac 1\Omega\int_{-\infty}^\infty d\epsilon f^{(L)}(\epsilon_+) \left[1-f^{(R)}(\epsilon_-)\right]$ are the bosonic form factors composed from electron distribution functions $f(\epsilon)$ in the left and right leads and $\epsilon_{\pm} = \epsilon \pm \Omega/2$.  These form-factors can be interpreted as the probability densities to find electron-hole pairs excited by the tunneling processes in the leads~\cite{beenakker2003proposal,chtchelkatchev2009hierarchical}. Namely, $n^{\LR}_{\Omega>0}$ describes the distribution of electron-hole pairs with the electron sitting at the left and the hole on the right leads while the distribution function $n^{\RL}_{\Omega>0}$ corresponds to the opposite situation, see Fig.~\ref{fig.sys}.

Using the distribution functions $f^{(L)}(\epsilon,T_{\ssL})=f_F(\epsilon-eV/2,T)$ and $f^{(R)}(\epsilon,T_{\ssR})=f_F(\epsilon+eV/2,T)$, with $T$ being the temperature of both leads, we find the explicit form of the form-factors
\begin{subequations}
\begin{eqnarray}
    n^{\LR}_{\Omega}&=&\frac {\Omega-eV}\Omega N_B(\Omega-eV,T),
    \\
    n^{\RL}_{\Omega}&=&\frac {\Omega+eV}\Omega N_B(\Omega+eV,T),
\end{eqnarray}
\end{subequations}
with $ N_B(\Omega,T)$ being the equilibrium Bose distribution function. The function $P(\Omega)$ in Eq.~(\ref{Ig}) is the probability density for a tunneling electron to exchange energy $\Omega$ with the effective bath of electron-hole pairs in the grains. A similar function appears in the transport theory of ultrasmall tunnel junctions interacting with an electromagnetic environment~\cite{devoret1992single,nazarov2009quantum}. In this theory the current can be reduced to a form identical to Eq.~\eqref{Ig} with $P(\Omega)$ describing the probability of energy exchange with an electromagnetic environment~\cite{chtchelkatchev2009hierarchical}. In our case the function $P(\Omega)$ in Eq.~(\ref{Ig}) has the form
\begin{equation}\label{Pg}
P(\Omega) = \int_{-\infty}^\infty \delta\left(\Omega+\sum_{j=1}^n\omega_j\right)\prod_{i=1}^n \frac{\omega_i d \omega_i}{2\pi (E_c^{(i)})^2} N^{(g_i)}_{\omega_i},
\end{equation}
with $N^{(g)}_{\omega} =\frac 1\omega\int_{-\infty}^\infty d\epsilon f^{(g)}(\epsilon_+) \left[1-f^{(g)}(\epsilon_-)\right]$ being
the effective distribution function of electron-hole pairs in grain $i$ and $f^{(g)}(\epsilon)$ being the electron distribution function in the same grain. The identity $N^{(g)}_{\omega}=-[1+N^{(g)}_{-\omega}]$ ensures that function $P(\Omega)$ is positively defined and for frequencies $|\Omega|\ll E_c^{(i)}$, $i=1,\ldots,n$, is properly normalized. We note, that all integrals over the frequencies $\omega$ in Eq.~\eqref{Pg} are quickly converging for energies much smaller than the charging energy $E_c^{(i)}$ [on frequency-scales on the order of voltage or grain temperature].

For positive frequencies, $\omega_i > 0$, the factor $N^{(g_i)}_{\omega_i}$ in Eq.~\eqref{Pg} describes the transfer of energy from an electron-hole pair decaying in the grain $i$ to the cotunneling electron. For negative frequencies, $\omega_i<0$, we rewrite the product $\omega_iN^{(g_i)}_{\omega_i}=(-\omega_i)(1+N^{(g_i)}_{-\omega_i})$, which describes the process of excitation of one electron-hole pair in grain $i$ with energy $|\omega_i|$ by the cotunneling electron.

\begin{figure}[tb]
  \centering
  \includegraphics[width=\columnwidth]{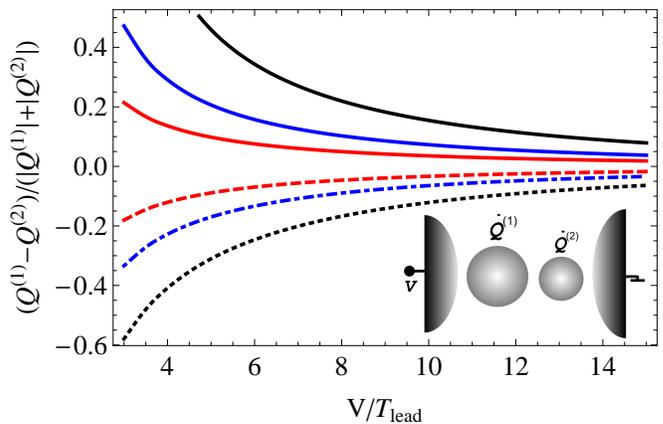}
  \\
  \caption{(Color online) The difference of the heat rates dissipated in the first and second grains of a two-grain junction shown
 in the inset with the different grain temperatures, $T_{g1}\neq T_{g2}$. The curves from top to bottom correspond to temperatures $T=T_{g1}=T_{\rm lead}$ and $T_{g2}/T=1.2,1.1,1.05,0.95,0.09,0.8$.}\label{fig3}
\end{figure}
Several methods exist to investigate the heat transfer in granular junctions especially for a single grain junction, when the heat flows from one lead to the other due to temperature gradients, see, e.g., Refs.~\cite{kindermann2004statistics,basko2005coulomb,giazotto2006opportunities,GlatzPhysRevB.81.033408.2010,glatz2010heating,laakso2010fully,heikkila2009statistics,kubala2008violation}.  Again, we follow the rate-approach to calculate the heat related to the excitation of electron-hole pairs by the cotunneling electron generalizing the method of Refs.~\cite{basko2005coulomb,chtchelkatchev2009hierarchical} and applying it to the case of arbitrary number of grains in the system. For the total electron-hole heat rate (the sum of the heat rates that electron donates to the electron-hole pairs in all grains in the course of inelastic cotunneling) we obtain
\begin{equation}\label{Qg}
\dot Q = \frac1{\mathcal R}\int_{-\infty}^\infty \Omega (n^{\LR}_{\Omega}+n^{\RL}_{\Omega}) P(\Omega)\Omega d\Omega\,.
\end{equation}
We note, that for leads (or/and the grains) with different temperatures there is an additional contribution to the heat, in addition to $\dot Q$, produced by temperature gradients: see, e.g., \cite{basko2005coulomb,kubala2008violation,laakso2010fully} and references therein.

Equation~\eqref{Qg} has an intuitive physical meaning if we introduce the bosonic form-factor $n_\Omega=\frac12(n^{\LR}_{\Omega}+n^{\RL}_{\Omega})$ describing the distribution of electron-hole pairs in the leads. This bosonic form-factor, contrary to $n^{\LR,\RL}_{\Omega}$, satisfies the relation $n_{-\Omega}=-[1+n_\Omega]$, which is the same as for Bose-functions. Taking into account the fact that the boson distribution function depends on positive frequencies we can rewrite Eq.~\eqref{Qg} in the form
\begin{equation}\label{Qg1}
\dot Q = \frac2{\mathcal R}\int_{0}^\infty \Omega [ n_{\Omega}P(\Omega)-(1+n_{\Omega})P(-\Omega)]\Omega d\Omega.
\end{equation}
The term $\Omega \, n_{\Omega}\, P(\Omega)$ in Eq.~(\ref{Qg1}) describes the amount of electron energy dissipated into the electron-hole bath inside the grains; the factor $n_{\Omega}$ implies an annihilation of the electron-hole pair in the leads. The term $\Omega(1+n_{\Omega})P(-\Omega)$ describes the amount of heat, which electron acquires from the electron-hole bath; $1+n_{\Omega}$ describes the creation of an electron-hole pair in the leads.

First we assume that all the grains and the leads have the same temperature $T$. In this case the distribution function
$N^{(g)}_{\omega}$ in Eq.~(\ref{Pg}) is the Bose function, $N_B(\omega) = 1/[\exp(\omega/T)-1]$ and the density distribution $P(\Omega)$ can be found analytically
\begin{equation}
\label{Pg2}
P(\Omega)= \frac{\Omega [1+N_B(\Omega)]}{(2\pi E_c^2)^n(2n-1)!}\prod_{j=1}^{n-1}\left[\Omega^2+(2\pi T j)^2\right],
\end{equation}
where $E_c\equiv (\prod_{i=1}^n E_c^{(i)})^{1/n}$.
From Eq.~\eqref{Pg2} follows that function $P(\Omega)$ satisfies the detailed balance symmetry relation, $P(-\Omega)=P(\Omega)\exp(-\Omega/T)$. In particular, for zero grain temperatures, $T=0$, the cotunneling electron can release its energy only into the grains, leading to $P(-\Omega) = 0$,~\cite{devoret1992single,nazarov2009quantum}.

Substituting Eq.~(\ref{Pg2}) for the $P(\Omega)$-function into Eqs.~\eqref{Ig} and \eqref{Qg} we find the current voltage characteristics $I(V, T)$ and the heat dissipated in the granular system, $\dot Q$, as follows
\begin{subequations}
\begin{equation}
\label{Ii}
I(V, T) = \frac{e^2V}{\mathcal R}\frac{\prod_{j=1}^{n}\left[(eV)^2+(2\pi T j)^2\right]}{(2n+1)! \, (2\pi E_c^2)^n},
\end{equation}
\begin{equation}
\label{Qq}
\dot Q = \frac{n}{n+1} \frac{(eV)^2}{\mathcal R}\frac{ \prod_{j=1}^{n}\left[(eV)^2+(2\pi T j)^2\right]}{(2n+1)!\, (2\pi E_c^2)^n}.
\end{equation}
\end{subequations}
Equation~\eqref{Ii} for the current is in agreement with the results of Refs.~\cite{devoret1992single,beloborodov2007granular}. The ratio of Eqs.~(\ref{Ii}) and (\ref{Qq})  reproduces our main result, Eq.~\eqref{eq1}, immediately.

\begin{figure}[tb]
  \centering
  \includegraphics[width=\columnwidth]{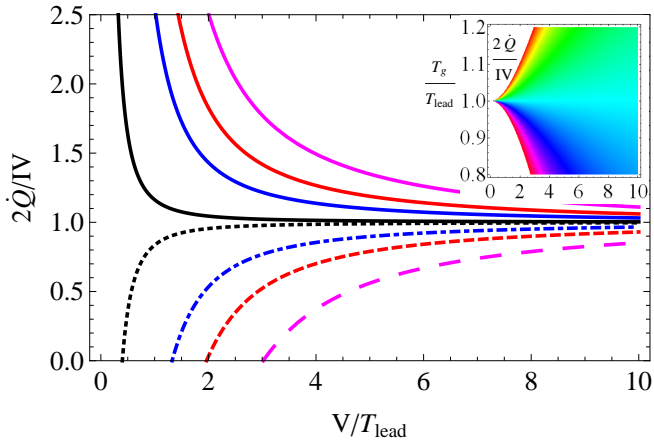}\\
  \caption{(Color online) The ratio of the heat $\dot Q$ dissipated in the grain to the total power $IV$, $2\dot Q/IV$ for a single grain junction
  with the grain temperature $T_g$ and the lead temperature $T_{\rm lead}$.  The curves from top to bottom correspond to $T_g/T_{\rm lead}=0.6,0.8,0.9,0.99,1.01,1.1,1.2,1.4$. The inset shows a density plot of $2\dot Q/IV$ as the function of $T_g/T_{\rm lead}$ and $V/T_{\rm lead}$ with $V$ being the bias voltage. }\label{fig2}
\end{figure}
Below we show that the amount of heat dissipated in
each grain is the same during the inelastic cotunneling process and it is given by the fraction $1/(n+1)$ of the total power $I\cdot V$.
Similar to Eq.~\eqref{Qg} we find the heat $\dot Q^{(i)}$ delivered by cotunneling electrons
to electron-hole pairs in grain $i$
\begin{equation}\label{Qi}
  \dot Q^{(i)} = -\frac2{\mathcal R}\int_{-\infty}^\infty\omega_i n_{\Omega}\delta(\Omega+\sum_{j=1}^n\omega_j)\Omega d\Omega\prod_{s=1}^n \frac{\omega_s d \omega_s}{2\pi (E_c^{(s)})^2} N^{(g_s)}_{\omega_s}.
\end{equation}
Here the summation over the grain index $i$ reproduces Eq.~(\ref{Qg}) for $\dot Q$. It follows from Eq.~(\ref{Qi}) that for equal distribution functions $N^{(g_s)}_{\omega_s}$ (all grain temperatures are the same) the heat dissipated in each grain $\dot Q^{(i)}$ does not depend on the grain index $i$ and it is equal to $\dot Q^{(i)}=\dot Q/n$. Below we investigate whether $\dot Q^{(i)}$ depends on the grain index $i$ for different grain temperatures.

For a two-grain junction, $n=2$, shown in the inset in Fig.~\ref{fig3} it follows that the heat dissipated in each grain is different for different grain temperatures, $\dot Q^{(1)}\neq\dot Q^{(2)}$. However, for large voltages, $|eV|\gg |T_{g1}-T_{g2}|$, as it follows from Fig.~\ref{fig3}, an uniform distribution of the heat over the grains is asymptotically rebuilt.

The fundamental question to address is the stability of the quantization condition \eqref{eq1} with respect to temperature
differences between grains and the leads. For a single grain junction, $n=1$, with lead and the grain
temperatures $T_{\rm lead}$ and $T_g$, respectively, we obtain, using Eqs.~\eqref{Ig}-\eqref{Pg} and~\eqref{Qg}
\begin{equation} \label{ratio1}
 \frac{\dot Q}{I \, V}= \frac12+\frac{T_{\rm lead}^2-T_g^2}{(eV)^2}\frac{(eV)^2+\frac85\bar T^2}{(eV)^2+(2\pi \bar T)^2}\pi^4,
\end{equation}
where $\bar T=\sqrt{(T_{\rm lead}^2+T_g^2)/2}$ is the characteristic temperature scale. For equal lead and grain temperatures, $T_{\rm lead}=T_g$,
the ratio in Eq.~\eqref{ratio1} is exactly one half, $\frac{\dot Q}{I \, V}= 1/2$, in accordance with Eq.~\eqref{eq1}.
For different temperatures, $T_{\rm lead}\neq T_g$, numerical calculations show that the quantization
of the ratio in Eq.~\eqref{ratio1} is smeared out, see Fig.~\ref{fig2}. However, even in this case for temperatures
$|T_{\rm lead}^2-T_g^2|/ (eV)^2\ll1$, as it follows from Eq.~\eqref{ratio1}, the ratio approaches one half,
see Fig.~\ref{fig2}.

We obtain qualitatively similar result for junctions with many grains, $n>1$. We define the vector of temperatures: $\vec \tau=(T_{\rm lead},T_g^{(1)},\ldots,T_g^{(n)})$, the average temperature $\bar \tau=\frac1{n+1}\sum \tau$, and the dispersion $\delta \tau=\sqrt{\frac1{n}\sum (\tau-\bar \tau)^2}$ and show that Eqs.~\eqref{eq1}-\eqref{eq2} are valid for voltages $|eV|/n \gg \delta \tau$.

Another important consequence of Eq.~\eqref{ratio1} is the appearance of a characteristic voltage scale $V_0(T_{\rm lead},T_g)$ for which $\dot Q=0$
\begin{equation}
\label{V0}
  V_0=\pi\sqrt2\left[\left(\frac{2 T_g^4 + 3 T_{\rm lead}^4}{5}\right)^{1/2}-T_{\rm lead}^2\right]^{1/2}.
\end{equation}
This voltage scale can be physically understood as follows: for bias voltages larger than $V_0$, $V > V_0$, the cotunneling electrons heat the grain, while for bias voltages smaller than $V_0$, $V< V_0$, the grain is cooling. It follows from Eq.~(\ref{V0}) that the voltage scale $V_0$ is nonzero, $V_0>0$, for grain temperatures larger than the lead temperatures, $T_g> T_{\rm lead}$ and it is zero, $V_0=0$, for equal temperatures $T_{\rm lead}=T_g=T$, meaning that the grain heats up for any bias voltage $V$. Similarly defined voltage
scales appear in multigrain circuits.

Finally, we address the question of existence of universality and quantization of the heat to power ratio in a
realistic experimental system. ``Hot'' electron-hole pairs generated in the grains by the inelastic cotunneling process drive electrons in the grains away from equilibrium state while electron-electron and electron-phonon interactions in the grains do the opposite: they thermalize electron distribution. Many experiments on granular systems in the inelastic cotunneling regime~\cite{beloborodov2007granular,Ovadia} show that the effective temperature approximation for electrons~\cite{giazotto2006opportunities} in the grains (Fermi distribution with effective temperature)
describes well the transport measurements in nearly whole range of the phonon bath temperatures $T_{\rm bath}$ and driving biases $V$ (except the case of ultralow bath temperatures and very high biases). In our consideration above we assume that the grain temperature is
the effective electron temperature,  $T_i(V)$, which is found by solving the heat balance equations. Experiments on variable range hoping in granular systems (special case of inelastic cotunneling~\cite{beloborodov2007granular}) show that at small bias the effective electron temperature is equal with high accuracy to the temperature of the phonon bath while overheating starts with
relatively high bias. In our case, the heat dissipated into particular grain $i$ is $\dot Q^{(i)}=\kappa_{\rm e-ph}\, (T_i^k-T_{\rm bath}^k)$, $i=1,\ldots, n$, with $\kappa_{\rm e-ph}$ being the electron-phonon interaction constant and $k$ being integer $4, 5$ or $6$ depending on the particular electron-phonon interaction model~\cite{giazotto2006opportunities,Ovadia,Altshuler}. For small bias, $\dot Q^{(i)}$ is also small leading to the condition, $T_i(V) \approx T_{\rm bath}$ for all grains, $i$. This is the regime of validity of Eqs.~\eqref{eq1}-\eqref{eq2}. For
a single grain, the condition $|T_g- T_{\rm bath}|\ll T_{\rm bath}$ implies that $V^2 (V^2+T_{\rm bath}^2)/(\mathcal R \, E_c^2  \, \kappa_{\rm e-ph} \, T_{\rm bath}^k)\ll1$.

\section{Discussion: heat/power quantization and symmetry of $P(\Omega)$}

The ``quantization'' of heat dissipation we are discussing above is formally related to the specific structure of the probability distribution function, $P(\Omega)$. We show that the general expressions for the current and heat in the multigranular junction are similar to that in the ultrasmall tunnel junction connected to the electromagnetic environment. The important question then arises: if the same quantization effect can be observed in ultrasmall tunnel junction and what symmetry of distribution function $P(\Omega)$ would then provide the quantization.

\subsection{Electromagnetic environment: single frequency mode. }

First we consider ultrasmall tunnel junction interacting with electromagnetic environment with the single mode. The probability that tunneling electron share the energy $\Omega$ to this environment is given by the expression~\cite{devoret1992single,nazarov2009quantum}
\begin{gather}
  P(\Omega)=\sum_{k=-\infty}^\infty p_k\delta(\Omega-k\omega_0),
  \\
  p_k=e^{-\rho\coth\frac{\beta\omega_0}2}I_k\left(\frac\rho{\sinh(\beta\omega_0/2)}\right)\exp(k\beta\omega_0/2),
\end{gather}
where $\rho>0$ is the interaction constant between tunneling electron and environment. It follows that $p_{-k}=p_k \exp(-k\beta\omega_0)$. Then

\begin{multline}
  \int_{-\infty}^\infty \Omega (n^{\LR}_{\Omega}-n^{\RL}_{\Omega}) P(\Omega)d\Omega=
  \\
  \sum_{k=-\infty}^\infty \frac{p_ke^{-k\beta\omega_0/2}(k\omega_0-V)}{\sinh[\beta(k\omega_0-V)/2]}\sinh(\beta V/2).
\end{multline}
Similarly we find:
\begin{multline}
  \int_{-\infty}^\infty \Omega^2 (n^{\LR}_{\Omega}+n^{\RL}_{\Omega}) P(\Omega)d\Omega=
  \\
  \sum_{k=-\infty}^\infty \frac{p_ke^{-k\beta\omega_0/2}k\omega_0(k\omega_0-V)}{\sinh[\beta(k\omega_0-V)/2]}\sinh(\beta V/2).
\end{multline}

\begin{figure}[t]
  \centering
  \includegraphics[width=0.95\columnwidth]{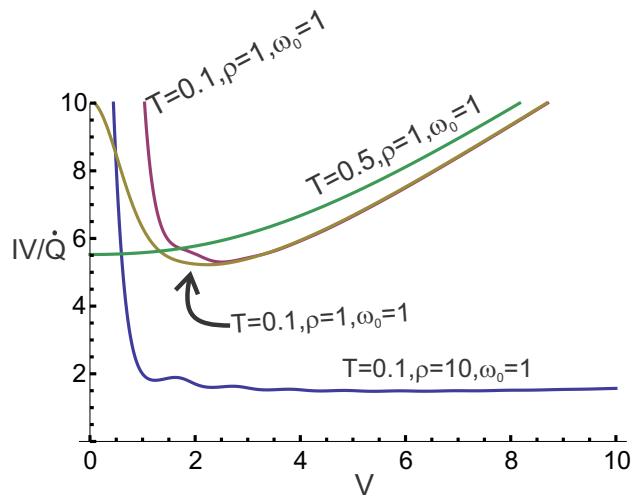}\\
  \caption{(Color online) The ratio $IV/\dot Q$ for  single mode environment.  This environment produces large dissipation when $V\sim\omega_0$. As follows, there is no universal quantization of the power to heat ration in general.}\label{fig1}
\end{figure}
Finally we obtain
\begin{gather}
 \frac {IV}{\dot Q}=\frac{V\sum\limits_{k=-\infty}^\infty I_k\left(\frac\rho{\sinh(\beta\omega_0/2)}\right)/\sinh[\beta(k\omega_0-V)/2]}{\sum\limits_{k=-\infty}^\infty k\omega_0 I_k\left(\frac\rho{\sinh(\beta\omega_0/2)}\right)/\sinh[\beta(k\omega_0-V)/2]}.
\end{gather}

The ratio $IV/\dot Q$ is shown in Fig.~\ref{fig1}. It follows that $IV/\dot Q$ is always far from unity. Therefore the environment absorbs only the part of the Joule heat as it is for the granular junction. However $IV/\dot Q$ does not quantize in general. Only in the limit $\rho\gg 1$ the ratio $IV/\dot Q$ starts approaching the integer number asymptotically. The question arises - why the ratio $IV/\dot Q$ does not quantize? We will address this question below after considering one more example.

\subsection{Electromagnetic environment: Ohmic impedance}

Now we consider the Ohmic environment related to the resistance embedded into the circuit~\cite{devoret1992single,nazarov2009quantum}. Without loss of generality we focus on the zero temperature limit, $T=0$. Then $P(\Omega)\propto\theta(\Omega)$ and we get
\begin{gather}
  \frac {IV}{\dot Q}=\frac{V\int_{0}^V (eV-\Omega)P(\Omega)d\Omega}{\int_{0}^V  \Omega (eV-\Omega)P(\Omega)d\Omega}.
\end{gather}

In this case for small frequencies (smaller than the charging energy $E_c$ of the tunnel junction) for distribution
function we have
\begin{gather}\label{P10}
  P(\Omega>0)\propto \Omega^{\frac2g-1},\qquad P(\Omega<0)=0,
\end{gather}
where $R$ is the resistance and $R_q$ is the resistance quantum and the parameter $g=\frac{R_q}R<1$.
Using Eqs.~(\ref{P10}) we obtain
\begin{gather}\label{11}
  \frac {IV}{\dot Q}=\frac{V\int_{0}^V (V-\Omega)P(\Omega)d\Omega}{\int_{0}^V  \Omega (V-\Omega)P(\Omega)d\Omega}=\frac{2 + 3 g +  g^2}{2 + g}=1+g.
\end{gather}
This result is valid for voltages below the charging energy, $V\ll E_c$. Equation~(\ref{11}) says
that the Ohmic electromagnetic environment is very effective as the absorber of energy. The universality of heat/power ratio does no take place.

\subsection{General $P(\Omega)$}
\begin{figure}
  \centering
  \includegraphics[width=\columnwidth]{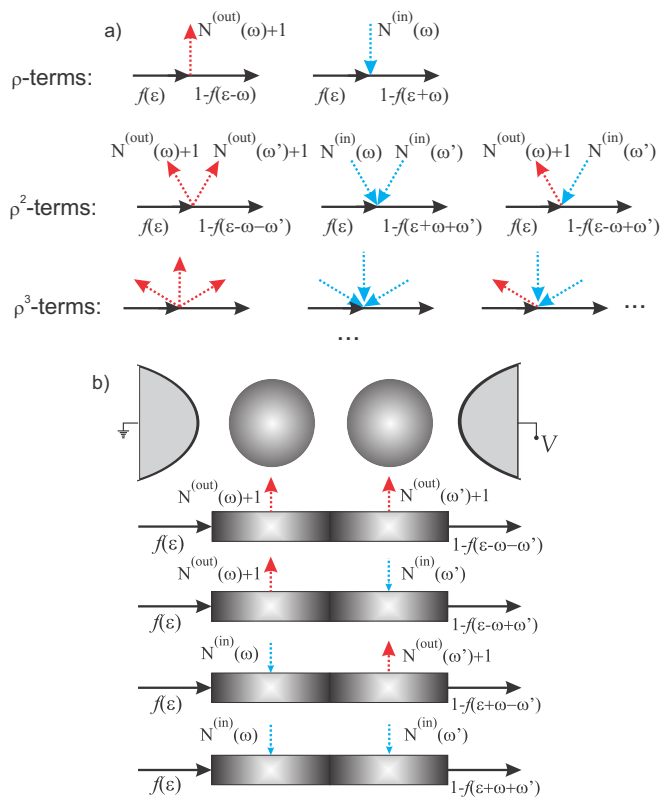}\\
  \caption{(Color online) a) The diagrams for $P(\Omega)$ in the first, second and partially in the third order over the electron-enevironment interaction parameter $\rho$ in the ultrasmall tunnel junction. b) All the diagrams for $P(\Omega)$ for the two-granular junction. The difference between the two cases is apparent from the symmetry structure of the diagrams. }\label{fig23}
\end{figure}
To summarise the results, we have demonstrated that in general there is no heat to power quantization in ultrasmall tunnel junctions while in the granular junction this quantization exists while they are in the cotunneling regime. The fundamental question is - what specific mechanism (``symmetry'') provides this quantization?  We understand the following: while electron goes through the granular junction it effectively produces the same number of electron-hole pairs in each grain (1/2 in leads) and this is the physical origin of the quantization. [In the sequantional tunneling regime there is no such environment of electron-hole pairs and so no universality.] This physics of the universality is hidden in the polynomial structure of the $P$-function prefactor as the function of $\Omega$ where the polynomial has the order of $2n-1$ over $\Omega$, with $n$ being the number of grains. A difficult question exists - what if we take an arbitrary polynomial of order $2n-1$ and build the $P$-function, would it produce the quantization effect? What mathematical restrictions should be given to the $P$-functions to provide the quantization? These important, however purely mathematical questions will be addressed in the forthcoming publication.

To distinguish the ``symmetry'' of the granular junction problem in the cotunneling regime from the ``symmetry'' of the ultrasmall tunnel junction problem in electromagnetic environment we show the diagrams that correspond the probability  $P(\Omega)$ in Fig.~\ref{fig23}. Here $N^{\rm (in)}$ and $N^{\rm (out)}$ are the Bose-functions. The symbols (in) and (out) here just help to understand what term effectively corresponds to emission or absorbtion of the environment (e-h) excitation. The notations we use here drawing these diagrams follow Ref.~\cite{chtchelkatchev2009hierarchical,Chtchelkatchev_Yalta2011}. In Ref.~\cite{chtchelkatchev2009hierarchical,Chtchelkatchev_Yalta2011} one can also find the detailed rules how to build $P(\Omega)$ from the diagrams. One should take the products of the distribution functions shown in the diagram and integrate over all frequencies with prime except $\omega$ to get the corresponding contribution to $P(\omega)$. So, here we demonstrate the difference of the topology (symmetry) of the diagrams for the ultrasmall tunnel junction problem and for the cotunneling problem in the granular junction.

\section{Conclusions}
In conclusion, we studied heating and dissipation effects in granular nanosystems in the regime of weak coupling between the grains. We focused on the cotunneling regime and solved the heat dissipation problem exactly in a chain of grains. We showed that, while the temperatures of the grains are kept equal, the power to heat ratio has an universal quantized value, Eq.~(\ref{eq1}), meaning that this ratio is geometrically  protected: it depends only on the number of grains. For different grain and lead temperatures the quantization effect is recovered asymptotically for large enough bias voltages.

\acknowledgments
N.~C. was supported by RFBR No.~13-02-00579, the Grant of President of Russian Federation for support of Leading Scientific Schools No.~6170.2012.2, RAS presidium and Russian Federal Government programs. A.~G. was supported by the U.S. Department of Energy Office of Science under the Contract No. DE-AC02-06CH11357.
I.~B. was supported by NSF Grant DMR 1158666.

\bibliography{bibliography_cotunneling}
\end{document}